\documentclass[final,onefignum,onetabnum,oneeqnum]{siamart171218}
\pdfoutput=1
\usepackage{lipsum}
\usepackage{amsfonts}
\usepackage{graphicx}
\usepackage{epstopdf}
\usepackage{algorithmic}
\ifpdf
\DeclareGraphicsExtensions{.eps,.pdf,.png,.jpg}
\else
\DeclareGraphicsExtensions{.eps}
\fi

\newsiamremark{remark}{Remark}
\newsiamremark{hypothesis}{Hypothesis}
\crefname{hypothesis}{Hypothesis}{Hypotheses}
\newsiamthm{claim}{Claim}

\headers{Riesz projection method for NLEVPs}{F. Binkowski, L. Zschiedrich, and S. Burger}

\title{A Riesz-projection-based method for \\ nonlinear eigenvalue problems}

\newcommand{\ZIB}{Zuse Institute Berlin, Takustra{\ss}e 7, 14195 Berlin, Germany}
\newcommand{\JCM}{JCMwave GmbH, Bolivarallee 22, 14050 Berlin, Germany}

\author{Felix Binkowski\thanks{\ZIB.}
	\and Lin Zschiedrich\thanks{\JCM.}
	\and Sven Burger\footnotemark[1]\hspace{0.11cm}\footnotemark[2]}

\usepackage{amsopn}

\usepackage{mathtools}
\usepackage[inline]{enumitem}

\ifpdf
\hypersetup{
  pdftitle={A Riesz-projection-based method for nonlinear eigenvalue problems},
  pdfauthor={F. Binkowski, L. Zschiedrich, and S. Burger}
}
\fi

\newcounter{simplecount}
\newcommand{\owncount}{\refstepcounter{simplecount}}

\begin{document}
\hyphenation{nano-disk}

\maketitle

\begin{abstract}
	We propose an algorithm for general nonlinear eigenvalue problems
	to compute physically relevant eigenvalues within a chosen contour. Eigenvalue information is
	explored by contour integration incorporating different weight functions.
	The gathered information is processed by solving a nonlinear system
	of equations of small dimension prioritizing eigenvalues
	with high physical impact. No auxiliary functions have to be introduced
	since linearization is not used. The numerical implementation
	is straightforward as the evaluation of the integrand
	can be regarded as a blackbox. We apply the method to
	a quantum mechanical problem and to two nanophotonic systems.
\end{abstract}


\section{Introduction}
\label{sec:Intro}
Nonlinear eigenvalue problems (NLEVPs) occur in many fields in physics,
from the dynamic analysis of macroscopic
structures~\cite{Tisseur_SIAMrev_2001}
to the investigation of photonic resonators on
the nanoscale~\cite{Novotny_Hecht_2012}, or scattering resonances
in quantum mechanics~\cite{Zworski_Scattering_Resonances_2019}.
The NLEVPs are solved numerically~\cite{Mehrmann_GAMM_2004,Guettel_NLEVP_2017}
in order to physically characterize
the systems~\cite{Taflove_Artech_2013,Lalanne_QNM_Benchmark_2018}.

We address the most general problem class of NLEVPs
\begin{align}
    T(\lambda)v = 0, \label{eq:NLEVP}
\end{align}
where $T(\lambda) \in \mathbb{C}^{n \times n}$ is the eigenvector residual
function, $\lambda \in \mathbb{C}$
is an eigenvalue, and $v \in \mathbb{C}^n$ is an eigenvector corresponding to $\lambda$.
In physics,
NLEVPs often have the form 
$ A(\lambda)v = \lambda B(\lambda)v,  $
which can be brought into the above form with
$T(\lambda) = A(\lambda) - \lambda B(\lambda)$.
In many applications, $T(\lambda)$ is a very large matrix, while just a few
eigenpairs $(\lambda,v)$ are responsible for the physical behavior of the described problem.

For rational residual functions $T(\lambda)$, the NLEVPs can be cast into a linear form,
so that standard approaches,
such as the Arnoldi or the Jacobi-Davidson method, are applicable.
Such a linearization introduces auxiliary functions and increases the dimension
of the problem~\cite{Guettel_NLEVP_2017,Yan_PRB_2018}.
Material dispersion is often significant in physical
systems and modeled by measured material data.
To apply the approach of linearization, material data
have to be fitted by rational functions~\cite{Garcia-Vergara_2017}
and numerical costs grow with the order and number of poles of the fit.
Note that the Arnoldi and the Jacobi-Davidson method have been adapted
to solve NLEVPs directly~\cite{Voss_nonlinArnoldi_2004,Voss_nonlinJacobiDavid_2007}.
\footnote{\newline This work has been published:\\
F. Binkowski et al., J. Comput. Phys. 419, 109678 (2020);
DOI: \href{https://doi.org/10.1016/j.jcp.2020.109678}{10.1016/j.jcp.2020.109678}}

In recent years, eigensolvers based on contour integration
attracted attention as they inherently support solving
NLEVPs~\cite{Asakura_JSIAM_2009,Beyn_LAAppl_2012,Yokota_JSIAM_2013,Gavin_JCompPhy_2018}.
Contour integral methods essentially involve solving 
linear systems of equations $T(\lambda)^{-1}y$ with random vectors $y$
along a chosen integration contour. The contour integration
gives a projection onto the eigenspace corresponding to the eigenvalues inside the contour.
In \cite{Yokota_JSIAM_2013,Gavin_JCompPhy_2018},
the Rayleigh-Ritz method is then used
to achieve approximations of eigenpairs.
The nonlinear structure is still inherited to the lower dimensional projected system
and needs to be given explicitly in a rational form. The methods of
\cite{Asakura_JSIAM_2009,Beyn_LAAppl_2012}
extract eigenvalue and eigenvector information by applying a singular-value decomposition
to the subspace generated by the contour integration and then solving a
linear eigenproblem.
An alternative way of extracting this information is based on
canonical polyadic tensor decomposition \cite{vanBarel_ContInt_JoCompApplMath_2016}.

The methods proposed in~\cite{Asakura_JSIAM_2009,Beyn_LAAppl_2012,Yokota_JSIAM_2013,Gavin_JCompPhy_2018}
yield eigenpairs whose eigenvalues
are located inside a specific region in the complex plane. This region is typically
chosen according to the underlying physical problem. However,
due to the numerical discretization, insignificant eigenvalues may occur
close to the physically relevant eigenvalues. A prominent example are eigenvalues
resulting from the truncation of open resonant systems~\cite{Yan_PRB_2018}.
Due to the application of random vectors for the contour integral methods,
in such cases, many eigenpairs have to be computed and they are not classified according to their physical relevance.

In this work, we present a contour integral method
which only projects onto the physically relevant eigenspaces.
This is done by the choice of the vector $y$ corresponding
to a physical source field which does not significantly couple to the
undesired eigenvectors.
These may arise due to the numerical discretization.
The insignificant eigenvalues are then filtered out by a fit to a nonlinear
model based on Cauchy's residue theorem.
Instead of computing individual eigenvectors,
spectral projections are calculated.
As linearization is circumvented in any stage of the procedure,
any material dispersion relation can be included.

\section{Riesz projection method for NLEVPs}
\label{sec:RP}
This section derives an approach to compute eigenvalues $\lambda$ fulfilling Eq.~(\ref{eq:NLEVP})
which are located inside a chosen contour and to compute
associated spectral projections, so-called Riesz projections.
To start with, notation and theoretical background on elements of
complex analysis are introduced \cite[Section 4.4]{Nolting_BookElectrodyn3_2016}.
We consider Eq.~(\ref{eq:NLEVP}) with a regular matrix
function $T: \Omega \rightarrow \mathbb{C}^{n \times n}$,
where $\Omega \subset \mathbb{C}$. 
Let $\mathcal{G}: \mathbb{C}^{n} \rightarrow \mathbb{C}$
be a meromorphic function and $y \in \mathbb{C}^n$ be a random vector.
Let $\Gamma_k \subset \Omega$
be a contour which  encloses a single eigenvalue $\lambda_k$ of the
residual function $T(\lambda)$ and on which the function
$\mathcal{G}\left(T(\lambda)^{-1}y\right)$ is holomorphic.
The eigenvalue $\lambda_k$ is a pole of $\mathcal{G}\left(T(\lambda)^{-1}y\right)$ and the pole
is assumed to be of order $p$.
Then, the Laurent series for $\mathcal{G}\left(T(\lambda)^{-1}y\right)$ about $\lambda_k$ is given by
\begin{align}
    \mathcal{G}\left(T(\lambda)^{-1}y\right) = \sum_{n=-p}^{\infty} a_n(\lambda-\lambda_k)^n,\hspace{0.5cm}
    a_{n}(\lambda_k) := \frac{1}{2 \pi i} \oint \limits_{\Gamma_k}
    \frac{\mathcal{G}\left(T(\xi)^{-1}y\right)}{(\xi-\lambda_k)^{n+1}}\,
    d\xi  \in \mathbb{C}. \label{eq:Laurent}
\end{align}
The coefficient $a_{-1}(\lambda_k)$ is the so-called residue
\begin{align}
    {\mathrm{Res}}_{\lambda_k}\left(\mathcal{G}\left(T(\lambda)^{-1}y\right)\right) =
    \frac{1}{2 \pi i} \oint \limits_{\Gamma_k} \mathcal{G}\left(T(\lambda)^{-1}y\right)
    \, d\lambda \label{eq:residuum}
\end{align}
of $\mathcal{G}\left(T(\lambda)^{-1}y\right)$ at $\lambda_k$.
\subsection{Sketch of the approach}
\label{sec:RP_sketch}
To show the idea of this work,
a simple eigenvalue $\lambda_k$ of $T(\lambda)$ is assumed, i.e., $\lambda_k$ is
a pole of $\mathcal{G}\left(T(\lambda)^{-1}y\right)$ and has the order $p=1$.
With the aim of extracting eigenvalue information from Eq.~(\ref{eq:residuum}),
the scalar function $f(\lambda) = \lambda$ is introduced.
Then, Cauchy's integral formula leads to 
\begin{align}
    {\mathrm{Res}}_{\lambda_k} \,\left(\lambda \mathcal{G}\left(T(\lambda)^{-1}y\right)\right) = 
    \frac{1}{2 \pi i} \oint \limits_{\Gamma_k} \lambda \mathcal{G}\left(T(\lambda)^{-1}y\right)
    \, d\lambda  \nonumber \\ = 
    \frac{1}{2 \pi i} \oint \limits_{\Gamma_k} \frac{ \lambda}{\lambda-\lambda_k}
    a_{-1}(\lambda_k)
    \, d\lambda  \nonumber \\ = 
    \lambda_k {\mathrm{Res}}_{\lambda_k}\,
    \left(\mathcal{G}\left(T(\lambda)^{-1}y\right)\right), \nonumber
\end{align}
where the regular part of the Laurent series vanishes and
only the principal part remains. Rearranging
yields the eigenvalue
\begin{align}
    \lambda_k = 
    \frac{{\mathrm{Res}}_{\lambda_k}\left(\lambda \mathcal{G}\left(T(\lambda)^{-1}y\right)\right)}
    {	{\mathrm{Res}}_{\lambda_k}\left(\mathcal{G}\left(T(\lambda)^{-1}y\right)\right)}\nonumber
\end{align}
inside of the contour $\Gamma_k$.  

\subsection{Generalized approach}
\label{sec:numeric}
The idea of the previous subsection can be generalized.
Firstly, we assume that the pole $\lambda_k$ has an order $p \geq 1$ and consider a
function $f: \Omega \rightarrow \mathbb{C}$ 
which is holomorphic on $\Gamma_k$ and inside of $\Gamma_k$ yielding
\begin{align}
    \frac{1}{2 \pi i} \oint \limits_{\Gamma_k} f(\lambda) \mathcal{G}\left(T(\lambda)^{-1}y\right)
    \, d\lambda  &=
    \sum_{n=-p}^{-1} a_{n}(\lambda_k)\frac{1}{2 \pi i} \oint \limits_{\Gamma_k}
    f(\lambda)(\lambda-\lambda_k)^{n}	\, d\lambda   \label{eq:one_pole} \\ &= 
    \sum_{n=-p}^{-1} a_{n}(\lambda_k) \frac{f(\lambda_k)^{(-n-1)}}{(-n-1)!}. \nonumber 
\end{align}
The coefficients $a_{-p}(\lambda_k),\dots,a_{-1}(\lambda_k)$ correspond to the
Laurent series for $\mathcal{G}\left(T(\lambda)^{-1}y\right)$ in Eq.~(\ref{eq:Laurent}).
Secondly, we choose a contour $\Gamma \subset \Omega$
on which $\mathcal{G}\left(T(\lambda)^{-1}y\right)$ is holomorphic and which encloses
finitely many poles $\lambda_1, \dots, \lambda_m$ of $\mathcal{G}\left(T(\lambda)^{-1}y\right)$.
Cauchy's residue theorem is used \cite{Nolting_BookElectrodyn3_2016}
so that Eq.~(\ref{eq:one_pole}) can be extended to
\begin{align}
    \frac{1}{2 \pi i} \oint \limits_{\Gamma} f(\lambda) \mathcal{G}\left(T(\lambda)^{-1}y\right)
    \, d\lambda   = \sum_{j=1}^{m}
    \sum_{n=-p}^{-1} a_{n}(\lambda_j) \frac{f(\lambda_j)^{(-n-1)}}{(-n-1)!},
\label{eq:residue_theorem_many_poles}
\end{align}
where $f(\lambda)$ has to be holomorphic on $\Gamma$ and inside of $\Gamma$.
Equation~(\ref{eq:residue_theorem_many_poles}) contains the
eigenvalues $\lambda_1, \dots, \lambda_m$ of the residual function $T(\lambda)$.
To explore the information for these  $m$ unknowns, we
introduce several \emph{weight functions} $f_1(\lambda), \dots, f_M(\lambda)$
and construct the nonlinear system of equations (NLSE)
\begin{align}
    \mu_k = F_k(\lambda_1,\dots,\lambda_m),
    \hspace{0.5cm} k=1,\dots,M, \label{eq:nonlin_sys}
\end{align}
where
\begin{align}
    \mu_k &\coloneqq \frac{1}{2 \pi i} \oint \limits_{\Gamma}
    f_k(\lambda)\mathcal{G}\left(T(\lambda)^{-1}y\right)
    \, d\lambda, \nonumber \\
    F_k(\lambda_1,\dots,\lambda_m) &=  \sum_{j=1}^{m}
    \sum_{n=-p}^{-1} a_{n}(\lambda_j) \frac{f_k(\lambda_j)^{(-n-1)}}{(-n-1)!}.
    \nonumber
\end{align}
Solving this NLSE yields the eigenvalues inside of the contour $\Gamma$.
The NLSE can
be solved with standard solvers, e.g, nonlinear system solvers based on least-square algorithms.
In this work, we use \emph{fsolve} from MATLAB. Note that the residues $a_{-1}(\lambda_k)$ are functions
depending on $\lambda_k$. Instead of
evaluating these functions, we regard them as unknowns themselves. We further note that the function
$\mathcal{G}\left(T(\lambda)^{-1}y\right)$ can be chosen according to the underlying physical
problem, e.g., it can be a physical observable. In this way, the numerical solution of $T(\lambda)^{-1}y$
is considered as a blackbox.

\subsection{Riesz projections with physical source fields}
The Riesz projector
\begin{align}
    P(T(\lambda),\Gamma) \coloneqq \frac{1}{2 \pi i} \oint \limits_{\Gamma} T(\lambda)^{-1}
    \, d\lambda \label{eq:RPs}
\end{align}
for a matrix function $T(\lambda)$ and a contour $\Gamma$ projects
vectors $y$ onto the eigenspace associated with
the eigenvalues inside of $\Gamma$ \cite{Hislop_BookSpecTheoSchroed_1996}.
In particular, for a contour $\Gamma_k$ enclosing
one simple eigenvalue $\lambda_k$ and a random vector $y$
as in Sec.~\ref{sec:RP_sketch},
the corresponding eigenvector is given by $v_k =P(T(\lambda),\Gamma_k) y$.

By choosing $y$ corresponding to a physical source field,
it is possible to distinguish between physically relevant and nonrelevant eigenvalues.
To illustrate this, a discretized partial differential equation
$T(\lambda)u = y$, where $y\in\mathbb{C}^n$
is a source field and $u\in\mathbb{C}^n$
is the solution of the problem, is regarded.
We consider an eigenvalue $\lambda_k$ to be insignificant if the Riesz projection
$P(T(\lambda),{\Gamma}_k)y$, where $\Gamma_k$ encloses only $\lambda_k$, is of the order of a given target accuracy.
If the physical source field $y$ is orthogonal to an eigenvector $v_k$, then $P(T(\lambda),{\Gamma}_k)y$ is of the order
of the discretization error.
However, for many problems with insignificant eigenvalues, 
$P(T(\lambda),{\Gamma}_k)y$ is larger than the discretization error 
as the physical source field $y$ has a small coupling to the corresponding eigenvector.
The target accuracy has to be chosen according to this.
Note that insignificant eigenvalues have a small influence
on the NLSE given by Eq.~(\ref{eq:nonlin_sys}) and are therefore filtered out
by solving the NLSE.

In this work, we aim at computing spectral projections instead of individual eigenvectors.
When an eigenvalue $\lambda_k$ is approximated by the NLSE, we define a 
further contour $\tilde{\Gamma}_k$ around this eigenvalue and  compute the Riesz projection
$P(T(\lambda),\tilde{\Gamma}_k)y$. This contour may include also insignificant eigenvalues
which are fully incorporated in the Riesz projection.
Note that $\tilde{\Gamma}_k$ could be chosen, so that clustered physically
relevant eigenvalues lie inside the contour. In this way,
the corresponding eigenvectors are treated as a single spectral projection~\cite{Zschiedrich_PRA_2018}.

\subsection{Algorithm}
To study physical systems where NLEVPs given by Eq.~(\ref{eq:NLEVP}) occur,
we propose Algorithm~\ref{alg:RP_method}. The algorithm can be sketched as follows.
A contour $\Gamma$ which encloses the eigenvalues of interest
has to be chosen (Step~1). 
Depending on the physical problem, a source field $y$ and a function $\mathcal{G}$
are defined (Step~2).
The contour integrals
in Eq.~(\ref{eq:nonlin_sys}) are computed 
with a suitable quadrature rule (Step~3-4). The evaluation of the integrand
at the integration points
essentially requires to solve linear systems of equations $T(\lambda)^{-1}y$.

The calculated integrals serve as the input data
for solving the NLSE given by Eq.~(\ref{eq:nonlin_sys}) (Step~5). The nonlinear model
$F_k(\lambda_1,\dots,\lambda_{{m}})$ is chosen based on the expected
order of the physically
relevant poles within $\Gamma$ and based on the expected number $m$ of poles.
The number $m$ can be estimated with physical a priori knowledge or, e.g., with Cauchy's argument
principle counting the zeros and poles of a meromorphic function.
If $m$ is greater than the number of physically
relevant poles within $\Gamma$, then the algorithm
returns relevant eigenvalues and also results which are insignificant.
Here, the used nonlinear solver exploits more degrees of freedom,
where some insignificant unknowns $\lambda_k$ and $a_{-1}(\lambda_k)$
have a very small influence on the physically relevant solutions. 
For example, an insignificant eigenvalue $\lambda_k$ could be close
to the integration contour and the corresponding residue $a_{-1}(\lambda_k)$
could be so small that the product of these two unknowns is of the order of the 
given target accuracy. On the other hand, if $m$ is smaller than the number
of physically relevant eigenvalues, then the nonlinear model
$F_k(\lambda_1,\dots,\lambda_{{m}})$ does not fit to the physical problem
and Algorithm~\ref{alg:RP_method} returns unsuitable results.
The error can be estimated by computing Riesz projections and the
algorithm has to be restarted with an increasing $m$.
To solve the NLSE, an initial guess is required. If no a priori information about
the eigenvalues is available, then randomly chosen numbers inside of the contour $\Gamma$
are a possible choice.

The results of solving the NLSE are approximations to eigenvalues
where eigenvalues with high physical impact regarding the source field $y$ are prioritized.
By defining contours $\tilde{\Gamma}_k$ around these eigenvalues (Step~6), 
Riesz projections $P(T(\lambda),\tilde{\Gamma}_k)y$ can be computed (Step~7).
Note that more than one eigenvalue can be inside of $\tilde{\Gamma}_k$.
The algorithm returns approximations $\lambda_1,\dots,\lambda_m$
and corresponding Riesz projections $P(T(\lambda),\tilde{\Gamma}_1)y,\dots,P(T(\lambda),\tilde{\Gamma}_m) y$ (Step~8).

\begin{algorithm}[t] 
	\caption{ for NLEVPs $T(\lambda)v = 0,
		\hspace{0.5cm} T(\lambda) \in \mathbb{C}^{n \times n},v \in \mathbb{C}^{n},
		\lambda \in \mathbb{C}$}
	\label{alg:RP_method} \owncount  
	\begin{algorithmic}[1]
		\STATE{Choose:
			\hspace{0.1cm} contour $\Gamma$, quadrature rule, integration points $\hat{\lambda}_1,\dots,\hat{\lambda}_N$, \\
			\hspace{1.51cm} size $M$ of NLSE, weight functions $f_1(\lambda), \dots, f_{M}(\lambda)$, \\
			\hspace{1.51cm} model $F_k(\lambda_1,\dots,\lambda_{{m}})$ with number of unknowns $m$, \\
			\hspace{1.51cm}	 initial guess for NLSE}	
		\STATE{Define: \hspace{.25cm} physical source field $y \in \mathbb{C}^n$,\\
			\hspace{1.52cm} function $\mathcal{G}: \mathbb{C}^n \rightarrow \mathbb{C}$},
		e.g., physical observable \vspace{0.1cm}
		\STATE{Solve: \hspace{0.43cm} linear systems
			$T(\hat{\lambda}_k)\hat{v}_k =  y, \hspace{0.5cm} k=1,\dots,N$} \vspace{0.1cm}
		\STATE{Compute:\hspace{0.0cm} 
			$\mu_k \coloneqq \frac{1}{2 \pi i} \oint_{\,\,\Gamma} f_k(\lambda)
			\mathcal{G}\left(T(\lambda)^{-1}y\right) \, d\lambda, \hspace{0.5cm} k=1,\dots,M$,
			\\ \hspace{1.53cm} using quadrature rule with $\hat{v}_1,\dots,\hat{v}_N$ \vspace{0.1cm}}	
		\STATE{Solve: \hspace{0.43cm} NLSE
			$\mu_k = F_k(\lambda_1,\dots,\lambda_{{m}}), \hspace{0.5cm} k=1,\dots,M$
			\vspace{0.1cm}}
		\STATE{Define: \hspace{0.27cm} contours $\tilde{\Gamma}_1,\dots,\tilde{\Gamma}_m$ enclosing
			eigenvalues $\lambda_1,\dots,\lambda_m$} \vspace{0.1cm}	
		\STATE{Compute:\hspace{0.0cm} 
			Riesz projections $P(T(\lambda),\tilde{\Gamma}_k) y
			\coloneqq \frac{1}{2 \pi i} \oint_{\,\,\tilde{\Gamma}_k} T(\lambda)^{-1} y \, d
			\lambda, \hspace{0.5cm} k=1,\dots,m$ }
		\vspace{-0.35cm}	
		\RETURN \hspace{0.18cm} approximate eigenvalues $\lambda_1,\dots,\lambda_m$, \\
		\hspace{1.53cm} Riesz projections $P(T(\lambda),\tilde{\Gamma}_1) y,
		\dots, P(T(\lambda),\tilde{\Gamma}_m) y$
	\end{algorithmic}
\end{algorithm}

Algorithm~\ref{alg:RP_method} is parallelizable on two levels. Firstly, the 
complete algorithm
can be performed for different contours $\Gamma$ simultaneously.
This can be useful if eigenvalues in different regions are of interest.
Secondly, solving the linear systems for
the numerical integration can be done in parallel.

\subsection{Numerical realization}
\label{sec:numer_real}
We realize the numerical integration in Algorithm~\ref{alg:RP_method} with
an $N$-point trapezoidal rule. The integration path $\Gamma$ is a
circular contour, which leads to exponential convergence \cite{Trefethen_SIAM_Trapz_2014}.
The equidistant integration points are given by
$\hat{\lambda}_k = \lambda_0 + r e^{2 \pi i k/N}$, $k = 1, \dots, N$,
where $\lambda_0$ and $r$ are the center and the radius of $\Gamma$, respectively.
Note that recently, 
rational filter functions for contour integral discretizations
have been designed using optimization techniques \cite{vanBarel_ratFilt_LinAlgAppl_2016}.
However, for the sake of simplicity, we use a trapezoidal rule.
To solve the linear systems of equations $T(\hat{\lambda}_k)^{-1}y$,
an LU decomposition can be used.
The LU decomposition needs not
to be updated at each integration point. Instead,
for sufficiently small changes in $\hat{\lambda}_k$, the LU decomposition
of a previous evaluation can be
used as a preconditioner for iterative solving.
This leads to a more efficient
numerical implementation.
	
In the following section, we consider physical examples
with simple eigenvalues, i.e., the nonlinear model in Eq.~(\ref{eq:nonlin_sys}) simplifies to
\begin{align}
	F_k(\lambda_1,\dots,\lambda_m) =  \sum_{j=1}^{m}
	a_{-1}(\lambda_j) f_k(\lambda_j).
	\nonumber
\end{align}
Equation~(\ref{eq:nonlin_sys}) is solved
with the nonlinear solver \emph{fsolve} from MATLAB. We regard
the residues $a_{-1}(\lambda_1),\dots,a_{-1}(\lambda_m)$ as unknown variables themselves
and set $M=2m$ to construct non-underdetermined NLSEs.
This handling of the residues allows for a simpler numerical realization.
For the weight functions, we choose the scaled polynomials
\begin{align}
	f_k(\lambda) = \left(\frac{\lambda - \lambda_0}{r}\right)^{k-1},
	\hspace{0.5cm} k = 1,\dots,M. \nonumber
\end{align}
Due to the fact that the weight functions are known and
that we treat the residues as unknowns,
also the Jacobians can be provided for the nonlinear solver. 
	
\section{Application of the method}
\label{sec:Appl}
We apply  Algorithm~\ref{alg:RP_method} to a quadratic NLEVP resulting from the
Schr\"odinger equation and to two rational NLEVPs resulting from Maxwell's equations.
These quantum mechanical and nanophotonic examples are open systems,
which are described by non-Hermitian operators.
In physics, the eigensolutions
of such problems are usually called
resonant states or quasinormal modes (QNMs)
\cite{Zworski_Resonances_1999,Lalanne_QNMReview_2018,Kristensen_QNM_2020}.
Material dispersion is omnipresent in
such systems and the physical understanding
of the resonance phenomena through numerical simulations
is an active research topic.
A common approach is a resonance expansion~\cite{Yan_PRB_2018,Ching_RevModPhys_1998}, where 
the solutions of the open systems are expanded into weighted sums of QNMs.
In this context, the coupling of a QNM to a source field is quantified by
the corresponding single expansion term.
Physically relevant QNMs lead to significant contributions in the resonance expansion.
	
The physical systems are numerically discretized
with the finite element method (FEM)~\cite{Monk_2003,Weiser_FEM_2016}.
We use the software package JCMsuite to discretize and to solve the nanophotonic problems in
Secs.~\ref{sec:Appl2} and~\ref{sec:Appl3}.
Note that also other numerical methods and implementations
could be used for applying Algorithm~\ref{alg:RP_method}.
	
\begin{figure}
		\centering
		{\includegraphics[width=0.85\textwidth]{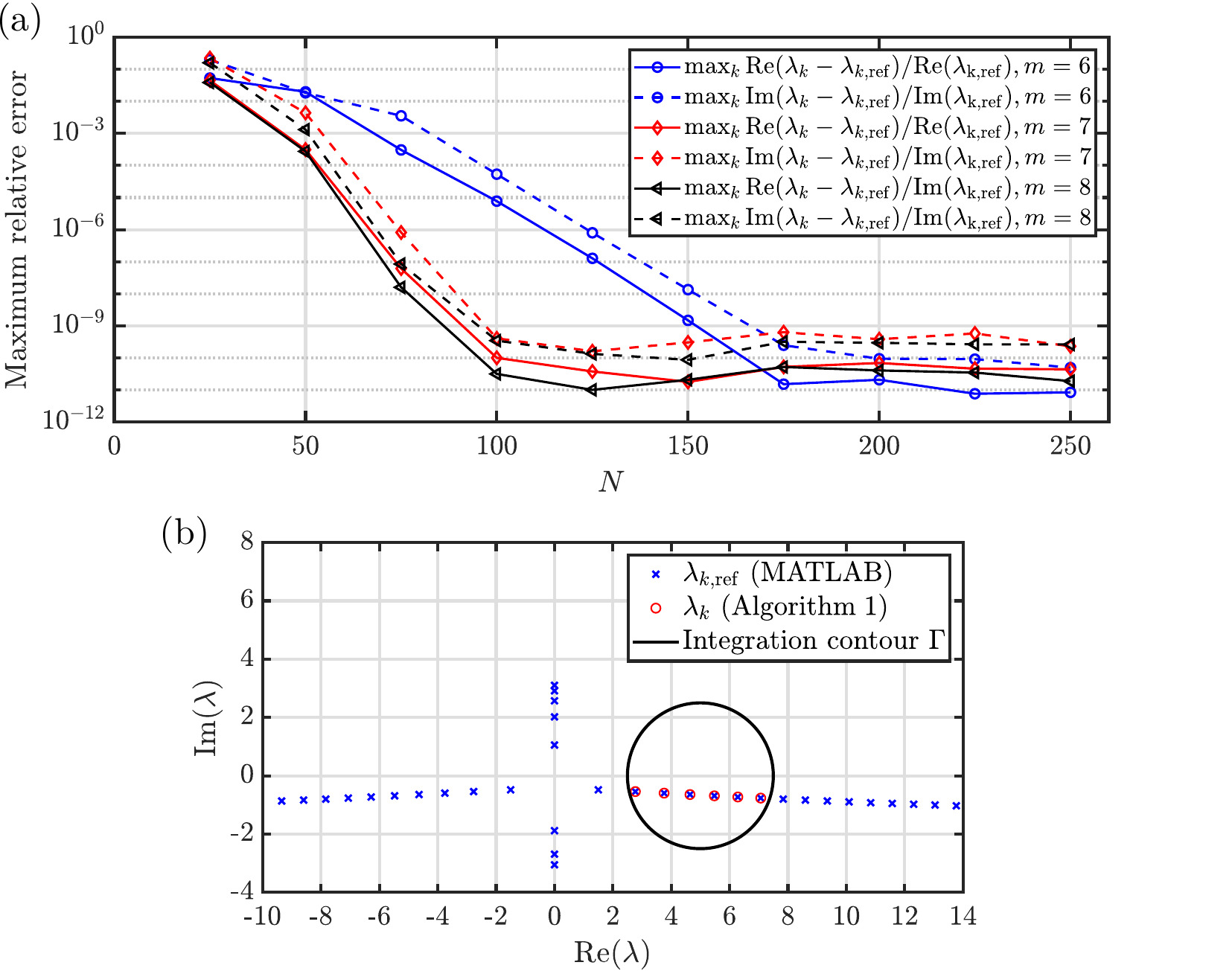}}
		\caption{\normalfont
			Results for computing eigenvalues of an
			open quantum system \cite{Shao_JapplPhys_1995}.
			(a)~Maximum relative errors,
			$\max_k \mathrm{Re}(\lambda_k-\lambda_{k,\mathrm{ref}})/\mathrm{Re}(\lambda_{k,\mathrm{ref}})$
			and 
			$\max_k \mathrm{Im}(\lambda_k-\lambda_{k,\mathrm{ref}})/\mathrm{Im}(\lambda_{k,\mathrm{ref}})$,
			$k=1,\dots,m$,
			as a function of  the number $N$ of integration points.		
			The eigenvalues $\lambda_k$ are computed with Algorithm~\ref{alg:RP_method}.
			The reference solutions $\lambda_{k,\mathrm{ref}}$ are computed
			by applying MATLAB's function \emph{eigs} to the
			linearized problem. (b)~Eigenvalues and integration contour. 
			The numerical integration is performed with $N=150$ integration points
			and the NLSE is solved for $m=6$ unknowns.	
		}
	\label{fig:fig01}
\end{figure}
	
\subsection{Resonant states in an open quantum system}
\label{sec:Appl1}
Propagation of a quantum particle of effective mass $m^*$
through a one-dimensional potential $V(x)$ can be described by
the time-independent Schr\"odinger equation 
\begin{align}
	-\frac{\hbar}{2} \frac{\partial}{\partial x}\left( \frac{1}{m^*}
	\frac{\partial \Psi(x)}{\partial x}  \right) + V(x) \Psi(x)
	= E \Psi(x), \label{eq:Schroedinger}
\end{align}
where $\hbar$ is the Planck constant, $E$ is the energy and $\Psi(x)$ is
the unknown wave function.
For a detailed description and motivation of
this example, see \cite{Shao_JapplPhys_1995}.
To compute the eigenvalues of this problem
in the domain $[-L,L]$, we use the approach of \cite{Gavin_JCompPhy_2018}
to scale and discretize Eq.~(\ref{eq:Schroedinger})
yielding the quadratic NLEVP 
\begin{align}
	T(\lambda)v = \left(\lambda^2A_2 + i\lambda A_1 - A_0\right)
	v = 0, \nonumber
\end{align}
	where
\begin{align}
	A_2 =
	\frac{h}{6}\begin{bmatrix}
	2 		& 1 	& 0 	  & 0 	  	 & \dots & 0 		\\
	1 		& 4 	& 1		  & 0 	  	 & \dots & 0 		\\
	0 		& 1 	& 4		  & 1 	  	 & \dots & 0 		\\
	\vdots 	&   	& 		  &\ddots	 &  	 & \vdots   \\
	0		&\dots 	& 0  	  & 1	  	 & 4	 & 1		\\
	0		&\dots 	& 0  	  & 0	 	 & 1	 & 2		
	\end{bmatrix}, \hspace{0.25cm}
	A_1 =
	\begin{bmatrix}
	1 		& 0 	  & \dots	& 0 	  	 & 0 		\\
	0 		& 0 		  & \dots 	& 0	  	 & 0		\\
	\vdots	& \vdots& \vdots  & \vdots	  	 & \vdots		\\
	0 		& 0 		  & \dots 	& 0	  	 & 0		\\
	0 		& 0 		  & \dots 	& 0	  	 & 1		\\	
	\end{bmatrix}, \nonumber
\end{align}
\begin{align}
	A_0 =
	\frac{1}{h}\begin{bmatrix}
	1 		& -1 	& 0 	& 0 	& \dots & 0 		\\
	-1 		& 2 	& -1	& 0 	& \dots & 0 		\\
	0 		& -1 	& 2		& -1 	& \dots & 0 		\\
	\vdots 	&   	& 		&\ddots	&  	 	& \vdots	\\
	0		&\dots 	& 0  	& -1	& 2	 	& -1		\\
	0		&\dots 	& 0  	& 0	 	& -1	& 1		
	\end{bmatrix} - V_0 A_2 \in \mathbb{R}^{n+2\times n+2}. \nonumber
\end{align}
The same
parameters as in \cite{Gavin_JCompPhy_2018} are chosen, where $L=\pi/ \sqrt{2}$,
$V(x)=V_0=10$,
and the spatial step size is $h = 2L/(n+1)$ with $n=302$.
	
In order to demonstrate
a numerical realization of Algorithm~\ref{alg:RP_method} which yields all eigenvalues inside a chosen contour,
we choose a unit random vector $y \in \mathbb{C}^{304}$ and the function
$\mathcal{G}(T(\lambda)^{-1}y) = x^HT(\lambda)^{-1}y$, where $x \in \mathbb{C}^{304}$
is also a unit random vector.
A circular contour $\Gamma$
with the center $\lambda_0=5$ and the radius $r=2.5$ is considered.
Solutions for an increasing number $N$ of integration points
and for different numbers of unknown eigenvalues, $m=6,\dots,8$, 
are computed. The maximum relative errors
$\max_k \mathrm{Re}(\lambda_k-\lambda_{k,\mathrm{ref}})/\mathrm{Re}(\lambda_{k,\mathrm{ref}})$
and 
$\max_k \mathrm{Im}(\lambda_k-\lambda_{k,\mathrm{ref}})/\mathrm{Im}(\lambda_{k,\mathrm{ref}})$
are shown in Fig.~\ref{fig:fig01}\hyperref[fig:fig01]{(a)}.
Exponential convergence up to a certain accuracy is obtained.
It can be further observed that for small $N$ and an increasing $m$,
the residuals become smaller. Here, the nonlinear solver can exploit more
degrees of freedom, i.e., also eigenvalues outside the contour can be approximated.
These solutions are discarded.
The results of Algorithm~\ref{alg:RP_method} can be compared with 
solutions of the linearized problem \cite{Gavin_JCompPhy_2018} computed with MATLAB's
function \emph{eigs}.
These reference solutions and the eigenvalues
computed by Algorithm~\ref{alg:RP_method} with $N=150$ and $m=6$  are
shown in Fig.~\ref{fig:fig01}\hyperref[fig:fig01]{(b)}.
Six eigenvalues inside of $\Gamma$ are obtained by
solving the linearized system.
The eigenvalues resulting from Algorithm~\ref{alg:RP_method}
coincide with these reference solutions.
	
\subsection{Photonic nanoantenna}
\label{sec:Appl2}
In the second numerical experiment,
we consider a nanophotonic structure.
Nanoantennas allow, e.g., for realizing single-photon emitters for
quantum technology devices \cite{Koenderink_SinglePhoton_2017}. 
We apply Algorithm~\ref{alg:RP_method} to an example from 
\cite{Zschiedrich_PRA_2018}, where a defect in a diamond nanodisk
is considered as solid-state single-photon emitter.
In the steady-state regime, the light-matter interaction of such a
structure can be described by the time-harmonic
Maxwell's equations in the second-order form
\begin{align}
	\nabla \times \mu(\mathbf{r},\omega)^{-1} \nabla \times \mathbf{E}(\mathbf{r},\omega) -
	\omega^2\epsilon(\mathbf{r},\omega) \mathbf{E}(\mathbf{r},\omega) =
	i\omega\mathbf{J}(\mathbf{r}), \label{eq:Maxwells_eq}
\end{align}
with the electric field $\mathbf{E}(\mathbf{r},\omega) \in \mathbb{C}^3$ and the source term
$\mathbf{J}(\mathbf{r}) \in \mathbb{C}^3$ as impressed current,
where $\mathbf{r} \in \mathbb{R}^3$ is the position.
The permittivity tensor $\epsilon(\mathbf{r},\omega)$
characterizes the spatial distribution of materials and, through its dependence on
the complex angular frequency $\omega \in \mathbb{C}$, the material dispersion. 
In the regime of optical frequencies, the permeability tensor $\mu(\mathbf{r},\omega)$
can typically be set to the vacuum permeability~$\mu_0$.
Equation~(\ref{eq:Maxwells_eq}) is discretized and solved with
the software package JCMsuite. The scattering solutions, i.e.,
solutions of Eq.~(\ref{eq:Maxwells_eq}) in presence of a source term,
are computed in the $\omega^2$ plane.
The outgoing radiation conditions for the diamond nanoresonator are realized with
perfectly matched layers (PMLs) \cite{Berenger_1994}.
We refer to \cite{Zschiedrich_PRA_2018}
for details on the FEM implementation and for details on the physical system.
	
\begin{figure}
		\centering
		{\includegraphics[width=0.85\textwidth]{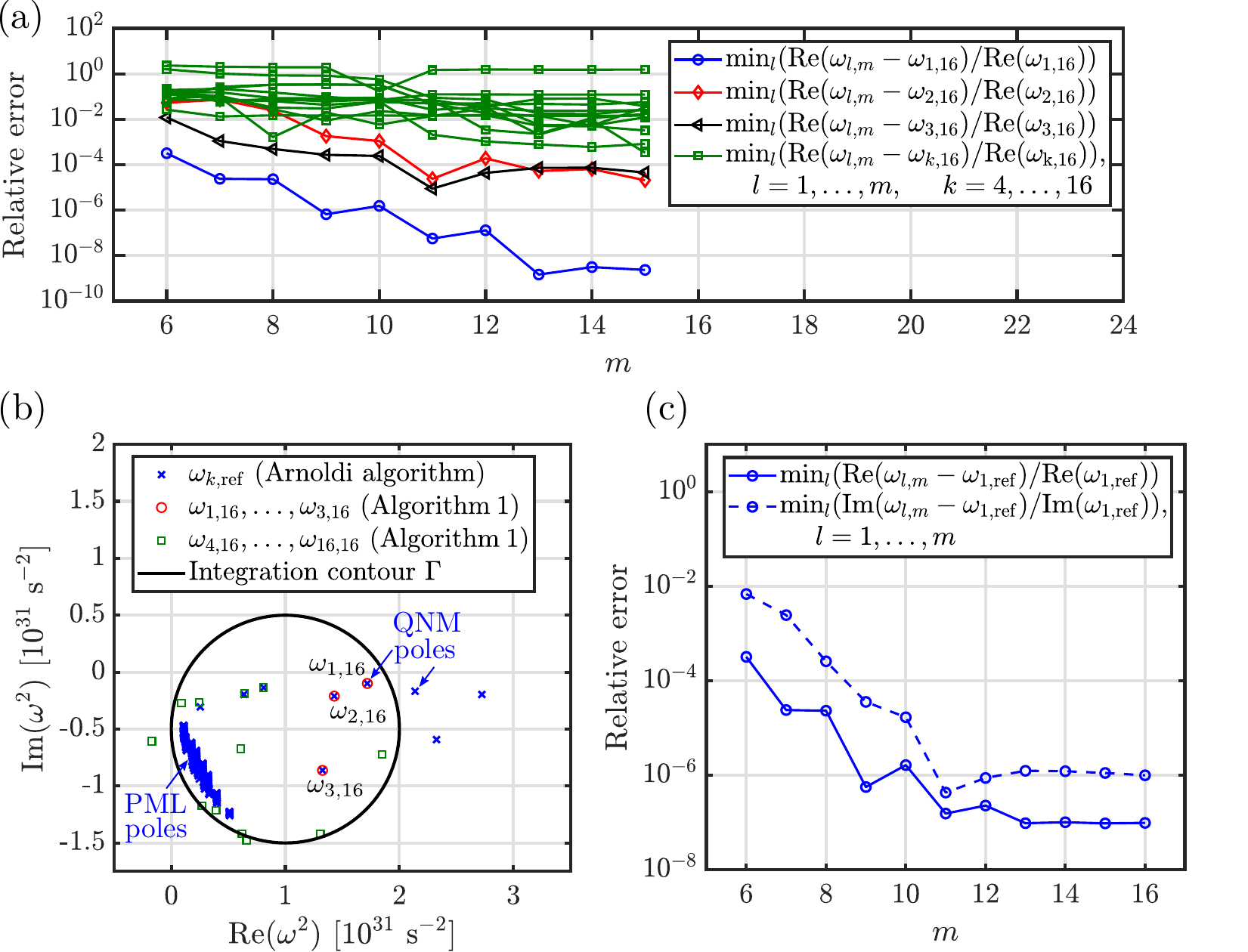}}
		\caption{\normalfont Results for computing
			eigenfrequencies of a dielectric nanoantenna \cite{Zschiedrich_PRA_2018}.
			Algorithm~\ref{alg:RP_method} leads to numerical convergence
			for the eigenfrequency $\omega_{1,16} = (1.7190-0.0992i)\times10^{31} \text{ s}^{-2}$.
			The numerical integration is performed using $N=200$ integration points.
			(a)	Relative errors for the real parts are shown
			as a function of the number of unknowns $m$ for the NLSE.
			The reference solutions $\omega_{k,16}$ are computed with $m=16$.	
			\mbox{(b)}~Eigenfrequencies $\omega_{k,16}$ computed with Algorithm~\ref{alg:RP_method}
			using $m=16$ unknowns.
			The reference solutions $\omega_{k,\mathrm{ref}}$ are computed with an Arnoldi algorithm.
			Physically relevant QNM poles are isolated eigenfrequencies and
			the physically nonrelevant PML poles lie in a cluster.	
			(e) Relative errors for the real and imaginary parts of the most accurate
			solution $\omega_{1,16}$,
			where the reference solution $\omega_{1,\mathrm{ref}}$ results from
			the Arnoldi algorithm.}
		\label{fig:fig02}
\end{figure}
	
The aim is to compute eigenfrequencies $\omega_k$ of the discretized NLEVP associated
with Eq.~(\ref{eq:Maxwells_eq}).
As the permittivity tensor $\epsilon(\mathbf{r},\omega)$
is a rational function, this is a rational NLEVP.
The interest is in eigenfrequencies
corresponding to physically relevant eigenvectors, the QNMs of the system, and, in particular,
in eigenfrequencies which QNMs couple to specific source fields. 
We use the source fields described in~\cite{Zschiedrich_PRA_2018}, which are
dipole emitters located inside the nanoresonator.
Accordingly, physical source fields $y$ for Algorithm~\ref{alg:RP_method} are chosen
as Eq.~(\ref{eq:Maxwells_eq}) is solved with dipole emitters for complex frequencies $\omega$ at 
the integration contour. The function $\mathcal{G}(T(\omega)^{-1}y)$ is a
point evaluation of the electric field at one chosen spatial point with a suitable scaling.
	
A circular contour $\Gamma$ is selected where the center is
$\omega_0^2 = (1 - 0.5i)\times10^{31} \text{ s}^{-2}$
and the radius is $r = 1\times10^{31} \text{ s}^{-2}$.
We apply Algorithm~\ref{alg:RP_method} while setting $N=200$ integration points
and solving the NLSE for $m=6,\dots,16$ unknowns.
The $l$-th solution of the algorithm is denoted
by $\omega_{l,m}$, where $l = 1,\dots,m$.
To study the numerical convergence of the real parts of the solutions, we use the
computed eigenfrequencies $\omega_{k,16}$, $k=1,\dots,16$, as reference solutions. 
Figure~\ref{fig:fig02}\hyperref[fig:fig02]{(a)} shows, 
for each $m=6,\dots,16$ and $k=1,\dots,16$, the minimum relative errors
$\mathrm{min}_l (\mathrm{Re}(\omega_{l,m}-\omega_{k,16})/\mathrm{Re(\omega_{k,16})})$,
where $l=1,\dots,m$. This means that at each step, the eigenfrequencies
which are closest to the reference solutions $\omega_{k,16}$ are identified.
Note that we investigate this error quantity
	because the aim is to show which of the solutions $\omega_{k,16}$, $k=1,\dots,16$, 
	are physically relevant. It is expected that physically relevant
	eigenfrequencies converge with respect to $m$ and, for insignificant eigenfrequencies,
	Algorithm~\ref{alg:RP_method} shows no convergence.
For the eigenfrequency
$\omega_{1,16} = (1.7190-0.0992i)\times10^{31} \text{ s}^{-2}$,
relative errors smaller than $10^{-8}$ are observed.
For the eigenfrequencies
$\omega_{2,16} = (1.429-0.211i)\times10^{31} \text{ s}^{-2}$ and
$\omega_{3,16} = (1.326-0.863i)\times10^{31} \text{ s}^{-2}$,
relative errors smaller than $10^{-4}$ are obtained. The results
$\omega_{4,16},\dots,\omega_{16,16}$ have relative errors larger than  $10^{-4}$.
	
Figure~\ref{fig:fig02}\hyperref[fig:fig02]{(b)}
shows the solutions $\omega_{k,16}$ using $m=16$
and solutions $\omega_{k,\mathrm{ref}}$ computed with the eigensolver from \mbox{JCMsuite},
which applies an Arnoldi algorithm using auxiliary fields.
Figure~\ref{fig:fig02}\hyperref[fig:fig02]{(c)} shows the
relative errors for the real and imaginary parts
of the solutions of Algorithm~\ref{alg:RP_method}
which are closest to the reference solution $\omega_{1,\mathrm{ref}}$
computed with the Arnoldi algorithm. For $m=16$,
we observe relative errors of about $10^{-7}$ and $10^{-6}$
for the real and imaginary part, respectively.
The accuracy limitation can be attributed to 
the accuracy of the scattering problem solver from JCMsuite.
	
The eigenvectors of the investigated NLEVP associated with Eq.~(\ref{eq:Maxwells_eq})
can be classified into QNMs and PML modes~\cite{Vial_PRA_2014,Yan_PRB_2018}.
The QNMs have a physical meaning and the corresponding eigenfrequencies, the QNM poles, 
are associated with the discrete spectrum of the underlying operator.
The PML modes correspond to eigenfrequencies which are associated with the continuous spectrum of
the operator. These PML poles are algebraic eigenfrequencies depending on the FEM discretization of
the open resonator system. As shown in~Fig.~\ref{fig:fig02}\hyperref[fig:fig02]{(b)},
the PML poles are clustered in the complex plane~\cite{Vial_PRA_2014}.
With this numerical experiment, we show that with Algorithm~\ref{alg:RP_method}
it is possible to compute a physically relevant eigenfrequency which
is close to insignificant eigenfrequencies.
The eigenfrequency $\omega_{1,16}$ corresponds
to the QNM which is responsible 
for the largest peak of the normalized dipole emission, the Purcell factor, in the frequency range 
$0.5\times10^{31} \text{ s}^{-2}\leq \mathrm{Re}(\omega^2) \leq 2.5\times10^{31} \text{ s}^{-2}$,
see~\cite{Zschiedrich_PRA_2018}. This means that this QNM has a significant
coupling to the used dipole sources
in this frequency range~\cite{Zschiedrich_PRA_2018}. The eigenfrequencies
$\omega_{2,16}$ and $\omega_{3,16}$ also correspond
to QNMs, however, the coupling is less significant. Algorithm~\ref{alg:RP_method}
does not converge to PML poles due to the small coupling of the PML modes
to the applied dipole sources.
	
\subsection{Resonances based on the hydrodynamic Drude model}
\label{sec:Appl3}
In the third numerical experiment, Algorithm~\ref{alg:RP_method} is applied to a nanostructure
described by the hydrodynamic Drude model. This material model takes spatially
nonlocal interactions of the electron gas and the light into account and is used for
describing light-matter interaction in
nanostructures on the scale of a few nanometers~\cite{Raza_2015,Schmitt_ComputPhys_2018}.
For nonmagnetic materials, the inclusion of nonlocal material properties leads to the
coupled system of equations
\begin{align}
	\nabla \times \mu_0^{-1} \nabla  \times \mathbf{E}(\mathbf{r},\omega)
	- \omega^2\epsilon_\mathrm{loc}(\mathbf{r},\omega) \mathbf{E}(\mathbf{r},\omega) 
	&= i \omega \mathbf{J}_{\mathrm{hd}}(\mathbf{r},\omega) + 
	i \omega \mathbf{J}(\mathbf{r},\omega), \label{eq:coupled_system_hydro1} \\
	\beta^2 \nabla \left( \nabla \cdot \mathbf{J}_{\mathrm{hd}}(\mathbf{r},\omega) \right)
	+\omega\left( \omega+i \gamma \right) \mathbf{J}_{\mathrm{hd}}(\mathbf{r},\omega) 
	&=	i \omega \omega_\mathrm{p}^2 \epsilon_0 \mathbf{E}(\mathbf{r},\omega) \label{eq:coupled_system_hydro2}
\end{align}
for the electric field $\mathbf{E}(\mathbf{r},\omega)$ and the hydrodynamic
current density $\mathbf{J}_{\mathrm{hd}}(\mathbf{r},\omega)$.
The current density $\mathbf{J}(\mathbf{r},\omega)$ is the impressed source field.
The permittivity tensor $\epsilon_\mathrm{loc}(\mathbf{r},\omega)$
corresponds to the local material response, $\epsilon_0$ is the 
vacuum permittivity, and $\mu_0$ is the vacuum permeability.
The plasma frequency $\omega_\mathrm{p}$ and the 
damping constant $\gamma$ are associated with the 
local Drude model 
$\epsilon_\mathrm{d}(\omega) = \epsilon_0(\epsilon_\infty - \omega_\mathrm{p}^2/(\omega^2 +i \gamma \omega))$,
where $\epsilon_\infty$ is the relative permittivity at infinity.
The system constant $\beta = \sqrt{3/5}\,v_\mathrm{F}$ includes the
Fermi velocity $v_\mathrm{F}$~\cite{Boardman_1982}.
The coupled system given by Eqs.~(\ref{eq:coupled_system_hydro1}) and~(\ref{eq:coupled_system_hydro2})
is discretized and solved with the software package JCMsuite.
	
\begin{figure}[]
		\centering
		{\includegraphics[width=0.85\textwidth]{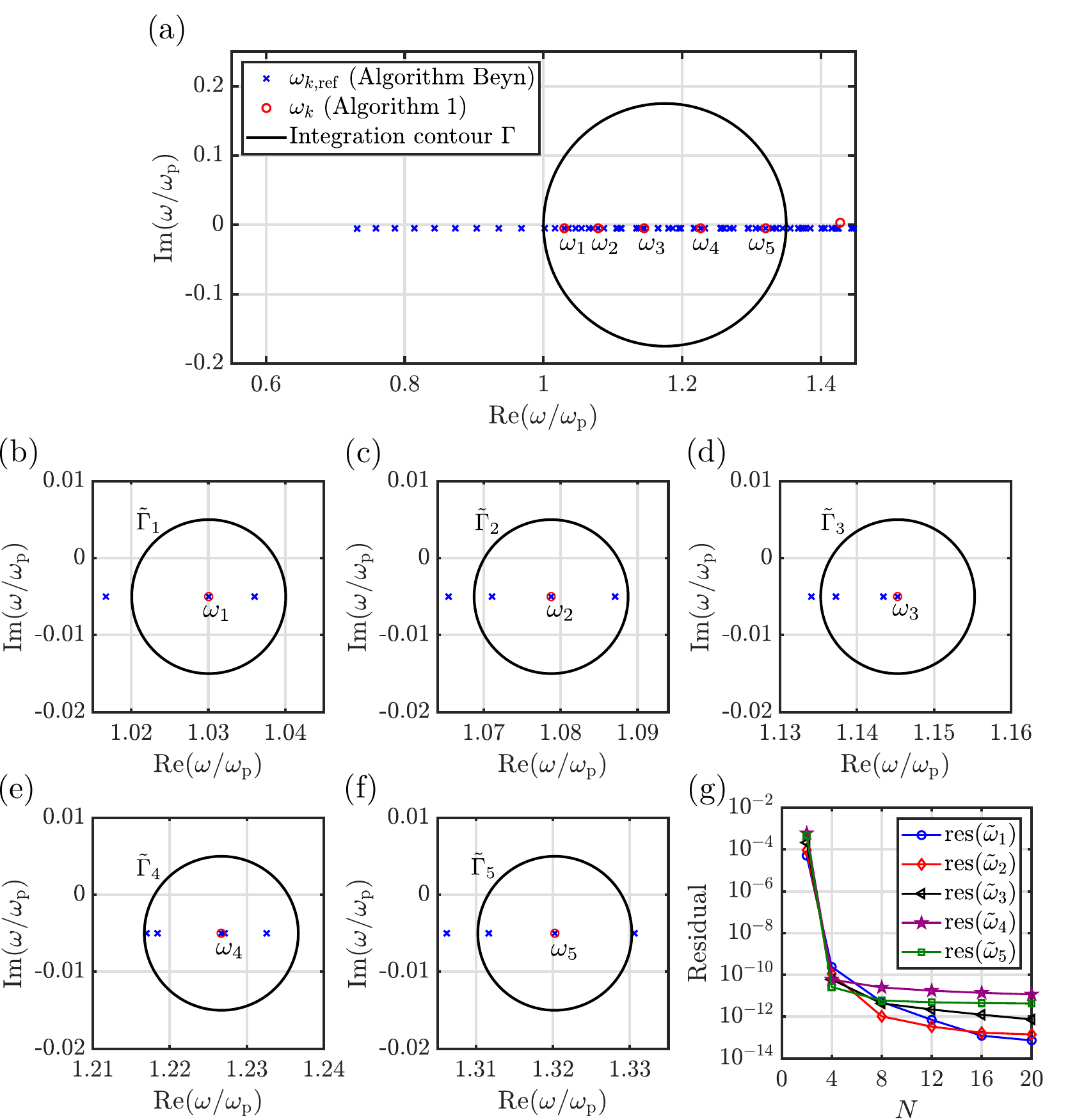}}
		\caption{\normalfont
			Results for computing
			eigenfrequencies and Riesz projections for a nanowire based on
			the hydrodynamic Drude model \cite{Binkowski_PRB_2019}.
			(a)~Algorithm~\ref{alg:RP_method} yields five eigenfrequencies,
			$\omega_1,\dots,\omega_5$, inside a chosen contour. The numerical
			integration is performed using $N=32$ integration points and the NLSE is solved for
			$m=6$ unknowns. The reference solutions $\omega_{k,\mathrm{ref}}$
			are taken from~\cite{Binkowski_PRB_2019},
			where an algorithm proposed by Beyn~\cite{Beyn_LAAppl_2012} is applied.	
			\mbox{(b-f)}~Integration contours $\tilde{\Gamma}_1,\dots,\tilde{\Gamma}_5$
			around the eigenfrequencies $\omega_1,\dots,\omega_5$.
			(g)~Residuals
			$\mathrm{res}(\tilde{\omega}_k)=||T(\tilde{\omega}_k)(P(T(\omega),\tilde{\Gamma}_k)y)||_2/||T(\tilde{\omega}_k)||_{F}$,
			where $||P(T(\omega),\tilde{\Gamma}_k)y||_2 = 1$,
			for each of the Riesz projections
			$P(T(\omega),\tilde{\Gamma}_1)y,\dots,P(T(\omega),\tilde{\Gamma}_5)y$	
			computed with a different number of integration points $N$.
			The eigenfrequencies $\tilde{\omega}_k$ are computed
			by Algorithm~\ref{alg:RP_method} for the contours $\tilde{\Gamma}_k$ with
			$m=1$ and the corresponding $N$.}
		\label{fig:fig03}
\end{figure}	
	
We consider a metal nanowire from~\cite{Binkowski_PRB_2019}, where
the eigenpairs have been computed using the contour integral
method proposed by Beyn~\cite{Beyn_LAAppl_2012}.
Furthermore, a modal analysis of the extinction cross section has been performed.
The nanowire has been illuminated by plane waves.
We refer to~\cite{Binkowski_PRB_2019} for details on the physical parameters
describing the nanowire and for details on the FEM realization.
Here, the aim is to compute only those eigenfrequencies which QNMs
have a significant coupling to these plane waves. 
Physical source fields $y$ for Algorithm~\ref{alg:RP_method} are chosen
by solving the coupled system given by Eqs.~(\ref{eq:coupled_system_hydro1})
and~(\ref{eq:coupled_system_hydro2}) for plane
waves with frequencies at the centers of the integration contours.
For all computations, we choose
$\mathcal{G}(T(\omega)^{-1}y) = \beta y^T T(\omega)^{-1}y$, where $\beta \in \mathbb{R}$ is
a scaling factor.
	
First, for the integration contour $\Gamma$ with a center at
$\omega_\mathrm{0} =1.175 \omega_p$ and a radius
of $r = 0.175\omega_p$, where $\omega_p = 8.65\times10^{15} \text{ s}^{-1}$,
Algorithm~\ref{alg:RP_method} is applied with $N=32$
integration points and $m=6$ unknowns.
Figure~\ref{fig:fig03}\hyperref[fig:fig03]{(a)} shows the results
of Algorithm~\ref{alg:RP_method} and reference solutions
from~\cite{Binkowski_PRB_2019}. Algorithm~\ref{alg:RP_method} yields five eigenfrequencies,
$\omega_1=(1.030-0.005i)\omega_\mathrm{p}$,
$\omega_2=(1.079-0.005i)\omega_\mathrm{p}$,
$\omega_3=(1.145-0.005i)\omega_\mathrm{p}$,
$\omega_4=(1.227-0.005i)\omega_\mathrm{p}$,
$\omega_5=(1.320-0.005i)\omega_\mathrm{p}$,
inside of $\Gamma$.
With respect to the reference solutions, the relative errors of their real and
imaginary parts are
smaller than $1.7 \times 10^{-5}$ and $1.5 \times 10^{-3}$, respectively. 
	
Secondly, the Riesz projections $P(T(\omega),\tilde{\Gamma}_k)y$ are
computed for contours $\tilde{\Gamma}_k$ around each of the eigenfrequencies
$\omega_1,\dots,\omega_5$. The contours 
$\tilde{\Gamma}_1,\dots,\tilde{\Gamma}_5$ are shown in
Fig.~\ref{fig:fig03}\hyperref[fig:fig03]{(b)}, \ref{fig:fig03}\hyperref[fig:fig03]{(c)},
\ref{fig:fig03}\hyperref[fig:fig03]{(d)}, \ref{fig:fig03}\hyperref[fig:fig03]{(e)},
and \ref{fig:fig03}\hyperref[fig:fig03]{(f)}, respectively.
Figure~\ref{fig:fig03}\hyperref[fig:fig03]{(g)} shows the residuals $\mathrm{res}(\tilde{\omega}_k) =||T(\tilde{\omega}_k)(P(T(\omega),\tilde{\Gamma}_k)y)||_2/||T(\tilde{\omega}_k)||_{F}$, where 
$||P(T(\omega),\tilde{\Gamma}_k)y||_2 = 1$,
for each of the Riesz projections computed with different numbers of integration points.
The eigenfrequencies $\tilde{\omega}_k$ are obtained by applying
Algorithm~\ref{alg:RP_method} for $\tilde{\Gamma}_k$ with different $N$
and a fixed $m=1$. The residuals become smaller with an increasing $N$.
For $N=20$, all residuals are smaller than $1.2 \times 10^{-11}$.
	
Algorithm~\ref{alg:RP_method} allows for computing only those
eigenfrequencies which QNMs couple to the plane wave
defined by the source field $y$.
The remaining eigenvectors which eigenfrequencies are located inside
the contour $\Gamma$ are also QNMs and they may are
relevant for another physical problem, however,
they are insignificant regarding
the here applied source field~\cite{Binkowski_PRB_2019}.
This means that small residuals can
be observed in Fig.~\ref{fig:fig03}\hyperref[fig:fig03]{(g)}
although the contours $\tilde{\Gamma}_k$ contain several eigenfrequencies.
The Riesz projections
$P(T(\omega),\tilde{\Gamma}_k)y$ are physically meaningful as they
mainly consists of contributions from the physically relevant QNMs.
In~\cite{Binkowski_PRB_2019}, the same eigenfrequencies have been identified as the 
physically relevant eigenfrequencies for the plane
wave excitation.
However, the approach from~\cite{Binkowski_PRB_2019}
requires calculation and investigation of the full electric fields corresponding
to the QNMs and is not as much straightforward.
	
\section{Conclusions}
\label{sec:Concl}
We presented a method based on
contour integration for computing eigenvalues and associated spectral
projections of general NLEVPs. Due to choosing specific physical source fields $y$
for the projection by contour integrals, only physically meaningful
eigenvalues are accessed. Instead of computing individual eigenvectors
corresponding to these eigenvalues, Riesz projections for frequency
ranges of interest are computed. In this way, an expensive computation
of a multitude of eigenpairs where most of them
are not physically relevant can be circumvented.
Numerical realizations were applied to non-Hermitian
problems from the fields of quantum mechanics and nanophotonics.
	
We considered the numerical solution of $T(\lambda)^{-1}y$ as 
a blackbox and extracted eigenvalue information by introducing
the meromorphic function $\mathcal{G}(T(\omega)^{-1}y)$, which can be a physical observable,
e.g., a point evaluation as in Sec.~\ref{sec:Appl2}.
Instead of global eigenfunctions, modal
contributions in form of Riesz projections, e.g., modal Purcell
factors \cite{Zschiedrich_PRA_2018} or modal extinction cross
sections~\cite{Binkowski_PRB_2019}, can be computed.
In this way, eigenvalues can still be extracted
without the need of a global approximation of the solution field $T(\lambda)^{-1}y$.
Therefore, we expect that  the algorithm will prove especially
useful for approaches without a vector representation of the solution field, such as
semi-analytical methods. Recently, the presented approach has been
compared with standard eigensolvers for NLEVPs resulting
from applications in nanophotonics~\cite{Lalanne_QNM_Benchmark_2018}. 
	
\section*{Acknowledgments}
We acknowledge support from Einstein Foundation Berlin within the framework
of MATHEON (ECMath project OT9)
and funding by the Deutsche Forschungsgemeinschaft
(DFG, German Research Foundation) under Germany's
Excellence Strategy -- The Berlin Mathematics Research
Center MATH+ (EXC-2046/1, project ID: 390685689, AA4-6).
We acknowledge the Helmholtz Association for funding within the Helmholtz Excellence Network
SOLARMATH, a strategic collaboration of the DFG Excellence Cluster
MATH+ and Helmholtz-Zentrum Berlin (grant no. ExNet-0042-Phase-2-3).

\end{document}